%% file: main.tex
\documentclass{article}

\usepackage[utf8]{inputenc}
\usepackage[T1]{fontenc}

\PassOptionsToPackage{table, dvipsnames}{xcolor}
\usepackage{xcolor}

\PassOptionsToPackage{colorlinks=true, citecolor=blue, linkcolor=blue, urlcolor=black}{hyperref}

\usepackage{arxiv}
\usepackage[utf8]{inputenc}
\usepackage[T1]{fontenc}
\usepackage{url}
\usepackage{booktabs}
\usepackage{nicefrac}
\usepackage{microtype}
\usepackage{lipsum}
\usepackage{graphicx}
\usepackage{natbib}
\usepackage{doi}
\usepackage{algorithm}
\usepackage{multirow}
\usepackage{multicol}
\usepackage{amsmath,amssymb,amsfonts}
\usepackage{amsthm}
\usepackage{mathrsfs}
\usepackage{float}
\usepackage{algorithmic}
\usepackage{bm}
\usepackage{caption}
\usepackage{hyperref}

\title{Meta-learning-enhanced implicit full waveform inversion
}

\author{
	{Zefeng~Wang, ~Weijian~Mao, ~Wei~Ouyang, ~Huanhuan~Tang} \\
	Research Center for Computational and Exploration Geophysics, \\State Key Laboratory of Precision Geodesy,\\
    Innovation Academy for Precision Measurement Science and Technology, \\ Chinese Academy of Sciences, Wuhan 430077, China \\
        \And
	\href{https://orcid.org/0000-0001-8868-7967}{\includegraphics[scale=0.06]{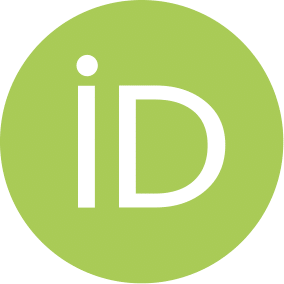}\hspace{1mm}Shijun~Cheng} \\
	Division of Physical Science and Engineering\\
	King Abdullah University of Science and Technology\\
	Thuwal 23955-6900, Saudi Arabia \\
    [3ex]
  $^{*}$Corresponding author: \textbf{Shijun Cheng}~(\texttt{sjcheng.academic@gmail.com})
}

\renewcommand{\shorttitle}{Meta-learning-enhanced implicit full waveform inversion}

\hypersetup{
pdftitle={A template for the arxiv style},
pdfsubject={q-bio.NC, q-bio.QM},
pdfauthor={David S.~Hippocampus, Elias D.~Striatum},
pdfkeywords={First keyword, Second keyword, More},
}

\begin{document}
\maketitle

\input{Sections/Abstract}
\input{Sections/Introduction}
\input{Sections/Method}
\input{Sections/Examples}
\input{Sections/Discussions}
\input{Sections/Conclusions}
\input{Sections/Acknowledgement}

\bibliographystyle{unsrtnat}
\bibliography{references}

\end{document}

%% file: Sections/Abstract.tex
\begin{abstract}
Implicit full waveform inversion (IFWI) introduces implicit neural representations to parameterize the subsurface velocity model as a continuous function of spatial coordinates, which alleviates the dependence on the initial model and improves inversion flexibility. However, IFWI still requires a large number of iterative updates for each new exploration area, leading to slow convergence, high computational cost, and a lack of mechanisms to share prior knowledge across different geological settings, thereby limiting its efficiency and generalization capability. To further accelerate convergence and enhance cross-area generalization, we propose a meta-learning-based implicit full waveform inversion method, referred to as Meta-learning-enhanced implicit full waveform inversion (Meta-IFWI). In this framework, the subsurface velocity model is represented using an implicit neural network with periodic activation functions (SIREN), while a meta-learning strategy is employed to pretrain a single network on multiple velocity inversion tasks. Through this process, the network learns shared inversion priors and rapid adaptation strategies across different geological scenarios. For a new inversion task, the pretrained Meta-IFWI model can be efficiently adapted to the observed seismic data with only a few gradient updates, significantly reducing the number of iterations required for inversion. Numerical experiments conducted on in-distribution models, including layered synthetic models and the Overthrust model, as well as out-of-distribution complex models such as Marmousi 2, demonstrate that, compared with conventional IFWI, the proposed Meta-IFWI achieves improved inversion accuracy while substantially accelerating convergence and reducing computational cost. Moreover, Meta-IFWI exhibits enhanced robustness and stronger cross-area generalization capability, validating its effectiveness and practical potential for seismic velocity inversion under complex geological conditions.
\end{abstract}

\keywords{Full-waveform inversion \and Meta-learning \and Implicit neural representation}

%% file: Sections/Introduction.tex
\section{Introduction}\label{sec1}
Modern hydrocarbon exploration increasingly targets structurally complex regions characterized by steep foreland thrust belts, intricate fault systems, and pronounced subsurface heterogeneity. In these settings, accurately recovering quantitative reservoir parameters (e.g., velocity and density) from seismic data is both critical and challenging. Full Waveform Inversion (FWI) serves as a core technique for high-resolution reconstruction of subsurface medium parameters, providing a powerful theoretical framework for seismic imaging and reservoir characterization by minimizing the mismatch between observed seismic data and numerically simulated wavefields \citep{tarantola1984inversion, virieux2010overview, pratt1999seismic}. However, traditional grid-based FWI methods still face several fundamental challenges in practice \citep{virieux1986p,levander1988}. First, FWI is highly nonlinear, and in the absence of low-frequency information or with an inaccurate initial model, the optimization process is prone to being trapped in local minima, leading to inversion failure or severe artifacts \citep{brossier2009seismic, tarantola1986strategy}. Second, the parameterization of the model is tightly coupled with spatial resolution: increasing resolution requires grid refinement, which substantially increases the computational burden of forward modeling and gradient calculation, while simultaneously exacerbating the ill-posedness of the optimization problem and hindering convergence \citep{gholami2013parameterization, xu20142d}. Third, as the complexity of exploration targets increases, conventional FWI exhibits severe limitations in achieving both high-accuracy and high-efficiency inversion in structurally complex regions \citep{prieux2011footprint, bian2010progress}. Together, these limitations motivate the search for different parameterization strategies.

One natural response has been to replace iterative inversion with data-driven approaches that directly learn the mapping from seismic observations to subsurface parameters \citep{lecun2015deep, yu2021deep, mousavi2024applications}. Feed-forward networks \citep{yang2019deep, wu2019inversionnet, yang2020deep, geng2022deep, du2022deep} take observed shot gathers as input and produce velocity or density models as output, entirely bypassing the forward-modeling loop and offering substantial speed advantages. Under favorable conditions, these methods achieve high-resolution reconstructions competitive with iterative FWI. However, their performance is dependent on the statistical match between training and target distributions. When geological conditions shift, generalization degrades markedly. Furthermore, physical constraints are typically imposed only implicitly through the training data, which limits the fidelity of predictions in structurally complex regions where training samples may be scarce. Finally, without explicit low-frequency content in the data, these networks struggle to recover large-scale velocity trends, a weakness inherited from the FWI problem itself \citep{yang2019deep}.

Implicit neural representations (INRs) offer a middle path between grid-based FWI and data-driven inversion. Rather than discretizing the velocity model onto a fixed grid or learning a dataset-level mapping, INRs use a neural network to represent velocity as a continuous function of spatial coordinates, with network weights serving as the inversion variables. This continuous parameterization decouples resolution from grid density, introduces implicit regularity through network architecture, and mitigates the ill-posedness inherent in conventional FWI. Architectures with periodic activations, such as SIREN \citep{sitzmann2020implicit}, are particularly effective in capturing fine-scale features. For example, \cite{sun2023implicit} introduced implicit neural representation (INR) into full waveform inversion, named IFWI, by parameterizing the velocity model with an SIREN network, replacing the conventional grid-based model parameters with network weights as the inversion variables, thereby leveraging the network's implicit regularization to improve inversion stability and mitigate ill-posedness. \cite{zhang2023multilayer} extended this concept to elastic FWI and demonstrated the effectiveness of INRs in handling complex inversion problems. \cite{yang2025gabor} combined Gabor wavelets with implicit neural representations for full waveform inversion, mitigating parameter crosstalk and enhancing the robustness of the inversion process. \cite{wang2025implicit} addressed the unbalanced gradient distribution in IFWI (caused by geometrical spreading) by isolating the gridded model gradients from the neural parameter gradients via the chain rule, and then applying a pseudo-Hessian to compensate for geometric spreading. These studies laid an important foundation for the development of IFWI. Collectively, these advances establish INR-based FWI as a viable and increasingly robust alternative to conventional methods. However, a significant practical limitation remains: each new survey area requires IFWI to be optimized from scratch, demanding a large number of iterations and offering limited transferability across geological settings.

Meta-learning, a machine learning paradigm designed to learn how to learn, trains models across a set of related tasks, enabling rapid adaptation to new tasks with only a small amount of task-specific data \citep{finn2017model, hospedales2021meta}. By leveraging a meta-trained model as a starting point, meta-learning can significantly reduce the time and computational resources required to achieve state-of-the-art performance on new tasks. Although its application in seismology remains relatively limited, existing studies have demonstrated the potential of meta-learning to enhance model adaptability and efficiency. For example, \cite{yuan2020adaptive} employed a meta-learning algorithm to address first-arrival picking across diverse seismic datasets, achieving higher accuracy than transfer learning approaches. In contrast, \cite{sun2020ml} applied meta-learning principles to develop an optimization strategy aimed at substantially accelerating the convergence of full waveform inversion. \cite{cheng2024meta} proposed a comprehensive framework, termed Meta-Processing, for a variety of seismic processing tasks. This approach leverages minimal training data for meta-learning, providing a general-purpose network initialization with strong adaptability. Moreover, \cite{cheng2025meta, cheng2026multifrequency} applied meta-learning to improve the efficiency and accuracy of neural network-based wavefield solutions, using a learned initialization strategy so that the network converges faster and generalizes better across different velocity models. These studies indicate that meta-learning offers significant advantages in reducing repetitive iterative optimization and improving cross-task adaptation efficiency, providing a direct and compelling theoretical foundation for integrating meta-learning into IFWI to reduce the computational burden associated with extensive iterations in new survey areas.

Motivated by these complementary strengths, this paper proposes Meta-IFWI, a meta-learning-enhanced implicit full waveform inversion framework. The key idea is to parameterize the velocity model using an SIREN-based INR and embed this representation within a model-agnostic meta-learning scheme. During meta-training, the network is optimized across a collection of velocity inversion tasks spanning diverse geological structures, enabling it to internalize cross-task geological priors and inversion patterns. When deployed on a new survey area, Meta-IFWI adapts to the observed seismic data within a small number of inversion iterations, achieving stable and accurate velocity reconstruction without requiring re-optimization from scratch. This design simultaneously reduces sensitivity to the initial model, accelerates convergence, and enhances generalization to previously unseen geological settings. As a result, it offers a new paradigm for efficient and robust FWI in structurally complex exploration environments.

%% file: Sections/Method.tex
\section{Method}\label{sec2}

This section presents the proposed Meta-IFWI framework. We first introduce IFWI, which parameterizes the subsurface velocity model as a continuous, differentiable function of spatial coordinates via an implicit neural representation. Building on this foundation, we then integrate a model-agnostic meta-learning (MAML) framework into IFWI, enabling the network to learn shared geological priors across inversion tasks and rapidly adapt to new surveys with only a small number of gradient updates.

\subsection{Implicit full-waveform inversion}
Traditional FWI employs a discretized, grid-based parameterization of the subsurface, in which inversion accuracy, memory footprint, and computational cost are all tightly coupled to the grid resolution. To overcome this limitation, implicit full waveform inversion (IFWI) replaces the discrete model grid with an INR, encoding subsurface parameters as a continuous, differentiable function of spatial coordinates. 

Formally, let $\mathbf{x} \in \Omega \subset \mathbb{R}^{d}$ denote the spatial coordinates and $m(\mathbf{x})$ represent the physical parameters of interest (e.g., P-wave velocity). The subsurface model is then parameterized by a neural network $N_{\theta}$ with learnable weights $\theta$:
\begin{equation}\label{eq1}
    m(\mathbf{x}) = N_{\theta}(\mathbf{x}).
\end{equation}
In practice, $N_{\theta}$ is typically realized as a multilayer perceptron (MLP) or, more commonly, an SIREN, whose periodic activations enable efficient capture of high-frequency spatial variations \citep{sitzmann2020implicit}. By mapping any queried coordinate $\mathbf{x}$ to the corresponding model value, this continuous representation fundamentally decouples the model resolution from the number of inversion parameters.

Given the implicitly represented model $m(\mathbf{x})$, seismic wave propagation is governed by the forward modeling operator $\mathcal{F}$, and the synthetic data recorded at the receivers are expressed as
\begin{equation}\label{eq2}
    d_{\mathrm{syn}} = \mathcal{F}(N_{\theta}(\mathbf{x}),\, s),
\end{equation}
where $s$ denotes the source signature, and $\mathcal{F}(\cdot)$ solves the wave equation for a given velocity model and source. The inversion objective is to find the network weights $\theta$ such that the synthetic data best match the observed data $d_{\mathrm{obs}}$. This is usually achieved by minimizing the $\ell_2$ data-misfit loss:
\begin{equation}\label{eq3}
    \mathcal{L}_{\mathrm{IFWI}}(\theta) 
    = \frac{1}{2}\left\| 
        \mathcal{F}(N_{\theta}(\mathbf{x}),\, s) - d_{\mathrm{obs}}
      \right\|_{2}^{2},
\end{equation}
where $\mathbf{x}$ is evaluated at a set of spatial query points (e.g., all grid points of the computational domain). Gradients of $\mathcal{L}_{\mathrm{IFWI}}$ with respect to $\theta$ are obtained by chaining the adjoint-state method through the forward operator $\mathcal{F}$ and backpropagation through $N_{\theta}$, enabling end-to-end optimization entirely within the implicit function space.

Compared with conventional FWI, IFWI offers three key advantages. First, the continuous parameterization completely decouples resolution from storage cost, i.e., the same network can be queried at arbitrary spatial locations without increasing the number of parameters. Second, the network architecture itself imposes an implicit smoothness prior that partially regularizes the inversion and reduces sensitivity to high-frequency noise. Third, IFWI does not require an explicit gridded velocity model as initialization and, thus, the inversion starts directly from the network weights, which are typically initialized randomly or by a simple prior.

Nevertheless, IFWI has notable limitations that motivate the method proposed in this work. Optimizing over the high-dimensional, highly nonlinear weight space of a neural network generally demands a large number of iterations to 
converge, increasing the per-survey computational cost relative to standard FWI. More critically, random weight initialization provides no geological structure or physical prior, so the optimization remains susceptible to poor convergence, particularly under complex geological conditions where the misfit landscape is highly non-convex. Both limitations point to the same underlying need: a principled strategy for initializing $\theta$ that encodes cross-task geological priors and accelerates convergence. We address this need in the following subsection through a meta-learning framework.

\begin{algorithm}
\caption{Meta-learning-enhanced implicit full-waveform inversion (Meta-IFWI)}
\label{alg:meta-ifwi}
\begin{algorithmic}[1]
\REQUIRE Training dataset 
         $\{(m^{(n)}, d_{\mathrm{obs}}^{(n)})\}_{n=1}^{N}$
\REQUIRE Forward modeling operator $\mathcal{F}(\cdot)$
\REQUIRE Source wavelet $s$
\REQUIRE Inner-loop learning rate $\alpha$, 
         outer-loop learning rate $\beta$
\REQUIRE Number of inner-loop update steps $K$, 
         number of task pairs per iteration $T_n$
\ENSURE  Meta-initialized network parameters $\theta_0$

\STATE Randomly initialize meta-parameters $\theta_0$

\WHILE{not converged}
    \STATE Sample $2T_n$ pairs from the training dataset and partition 
           into support set 
           $\mathcal{S} = \{(m^{(i)}, d_{\mathrm{obs}}^{(i)})\}_{i=1}^{T_n}$
           and query set 
           $\mathcal{Q} = \{(m^{(i)}, d_{\mathrm{obs,q}}^{(i)})\}_{i=1}^{T_n}$
    \STATE Initialize accumulated meta-gradient: 
           $g_{\mathrm{meta}} \gets \mathbf{0}$

    \FOR{$i = 1$ \TO $T_n$}
        \STATE \textit{// Inner loop: adapt to support task $i$}
        \STATE Initialize task-specific parameters: 
               $\theta_0^{(i)} \gets \theta_0$
        \FOR{$k = 1$ \TO $K$}
            \STATE Query implicit velocity model: \\ 
                   \hspace{1em} $m_k^{(i)}(\mathbf{x}) 
                   = N_{\theta_{k-1}^{(i)}}(\mathbf{x})$
            \STATE Perform forward modeling on support data: \\
                   \hspace{1em} $d_{\mathrm{syn}}^{(i)} 
                   = \mathcal{F}(m_k^{(i)},\, s)$
            \STATE Compute support-set loss: \\
                   \hspace{1em} $\mathcal{L}_{\mathrm{FWI}}^{(i)}
                   (\theta_{k-1}^{(i)})
                   = \dfrac{1}{2}
                   \left\| d_{\mathrm{syn}}^{(i)} 
                   - d_{\mathrm{obs}}^{(i)} \right\|_2^2$
            \STATE Update task parameters: \\
                   \hspace{1em} $\theta_k^{(i)} = \theta_{k-1}^{(i)}
                   - \alpha\,\nabla_{\theta}\,
                   \mathcal{L}_{\mathrm{FWI}}^{(i)}(\theta_{k-1}^{(i)})$
        \ENDFOR

        \STATE \textit{// Outer loop: evaluate adapted parameters on 
               paired query task $i$}
        \STATE Query implicit velocity model with adapted parameters: \\
               \hspace{1em} $m_K^{(i)}(\mathbf{x}) 
               = N_{\theta_K^{(i)}}(\mathbf{x})$
        \STATE Perform forward modeling on query data: \\
               \hspace{1em} $d_{\mathrm{syn,q}}^{(i)} 
               = \mathcal{F}(m_K^{(i)},\, s)$
        \STATE Compute query-set loss: \\
               \hspace{1em} $\mathcal{L}_{\mathrm{FWI}}^{(i)}
               (\theta_K^{(i)};\, d_{\mathrm{obs,q}}^{(i)})
               = \dfrac{1}{2}
               \left\| d_{\mathrm{syn,q}}^{(i)} 
               - d_{\mathrm{obs,q}}^{(i)} \right\|_2^2$
        \STATE Accumulate meta-gradient: \\
               \hspace{1em} $g_{\mathrm{meta}} \gets g_{\mathrm{meta}} 
               + \nabla_{\theta_0}\,
               \mathcal{L}_{\mathrm{FWI}}^{(i)}
               (\theta_K^{(i)};\, d_{\mathrm{obs,q}}^{(i)})$
    \ENDFOR

    \STATE \textit{// Meta-update: one update after all $T_n$ pairs 
           are processed}
    \STATE Update meta-initialization: \\
           \hspace{1em} $\theta_0 \leftarrow \theta_0 
           - \beta\, g_{\mathrm{meta}}$
\ENDWHILE

\RETURN Meta-trained initialization $\theta_0$
\end{algorithmic}
\end{algorithm}

\subsection{Meta-learning-enhanced implicit full waveform inversion}
As discussed in the preceding subsection, a key limitation of IFWI is that it must be optimized from scratch for each new inversion task, i.e., starting from a randomly initialized $\theta$ with no geological prior and requiring a large number of forward simulations and gradient updates to converge. This per-inversion cost becomes prohibitive in multi-model or multi-scenario applications and motivates a principled strategy for learning a task-agnostic initialization.

To address this, we integrate an MAML 
framework \citep{finn2017model} into IFWI, yielding the proposed 
\textbf{Meta-IFWI}. The central idea is to jointly train a shared 
meta-initialization $\theta_0$ across a distribution of seismic inversion 
tasks, so that $\theta_0$ encodes geological structures and physical priors common across tasks. When deployed on a new inversion task, only a small number of gradient updates starting from $\theta_0$ are needed to obtain stable, high-accuracy velocity reconstructions and, thus, significantly reduce both computational cost and sensitivity to initialization. The complete meta-training procedure is summarized in Algorithm~\ref{alg:meta-ifwi}.

Let the full training dataset consist of $N$ pairs of subsurface velocity 
models and their corresponding observed seismic data 
$\{(m^{(n)},\, d_{\mathrm{obs}}^{(n)})\}_{n=1}^{N}$. At the beginning of 
each meta-training iteration, $2T_n$ pairs are randomly sampled from this 
dataset and partitioned into two disjoint subsets of equal size: a 
\textit{support set} 
$\mathcal{S} = \{(m^{(i)},\, d_{\mathrm{obs}}^{(i)})\}_{i=1}^{T_n}$ used 
to drive the inner-loop adaptation, and a \textit{query set} 
$\mathcal{Q} = \{(m^{(j)},\, d_{\mathrm{obs}}^{(j)})\}_{j=1}^{T_n}$ used 
to evaluate the meta-objective in the outer loop. The value of $T_n$ is 
constrained by GPU memory, since the second-order gradients required by MAML must be retained simultaneously for all $T_n$ inner-loop computations during the outer-loop update. 

The task-level IFWI loss for the $i$-th support task follows the same form as Equation~\eqref{eq3}:
\begin{equation}\label{eq4}
    \mathcal{L}_{\mathrm{FWI}}^{(i)}(\theta)
    = \frac{1}{2}\left\|
        \mathcal{F}(N_{\theta}(\mathbf{x}),\,s) - d_{\mathrm{obs}}^{(i)}
      \right\|_2^2.
\end{equation}
For each support task $i \in \{1, \ldots, T_n\}$, the inner loop starts 
from the shared meta-initialization $\theta_0$ and performs $K$ gradient 
descent steps using the support data $d_{\mathrm{obs}}^{(i)}$. At the 
$k$-th step, the implicit velocity model is queried as
\begin{equation}\label{eq5}
    m_k^{(i)}(\mathbf{x}) = N_{\theta_{k-1}^{(i)}}(\mathbf{x}),
\end{equation}
forward modeling is then applied to obtain the corresponding synthetic 
seismic data
\begin{equation}\label{eq6}
    d_{\mathrm{syn}}^{(i)} = \mathcal{F}\!\left(m_k^{(i)},\,s\right),
\end{equation}
and the parameters are updated via gradient descent on the task-level loss:
\begin{equation}\label{eq7}
    \theta_k^{(i)} = \theta_{k-1}^{(i)} 
    - \alpha\,\nabla_{\theta}\,\mathcal{L}_{\mathrm{FWI}}^{(i)}
      \!\left(\theta_{k-1}^{(i)}\right),
    \quad k = 1, \ldots, K,
\end{equation}
where $\alpha$ is the inner-loop learning rate and $\theta_0^{(i)} \equiv \theta_0$ for all tasks. After $K$ steps, 
$\theta_K^{(i)}$ constitutes the task-adapted parameters for support 
task $i$.

Once all $T_n$ inner-loop adaptations are complete, the outer loop evaluates the generalization of each adapted $\theta_K^{(i)}$ on its paired query task from $\mathcal{Q}$. Specifically, the $i$-th adapted parameters $\theta_K^{(i)}$, which are obtained by adapting to support task $i$, are used to compute the inversion loss on query task $i$, with observed data $d_{\mathrm{obs,q}}^{(i)}$. The meta-objective accumulates these query-set losses across all $T_n$ task pairs:
\begin{equation}\label{eq8}
    \mathcal{L}_{\mathrm{meta}}(\theta_0)
    = \sum_{i=1}^{T_n}\,\mathcal{L}_{\mathrm{FWI}}^{(i)}
      \!\left(\theta_K^{(i)};\, d_{\mathrm{obs,q}}^{(i)}\right).
\end{equation}
Crucially, the gradient $\nabla_{\theta_0}\mathcal{L}_{\mathrm{meta}}$ is 
accumulated across all $T_n$ task pairs before any update to $\theta_0$ is performed. Once all contributions are aggregated, a single meta-update is applied:
\begin{equation}\label{eq9}
    \theta_0 \leftarrow \theta_0 
    - \beta\,\nabla_{\theta_0}\,\mathcal{L}_{\mathrm{meta}}(\theta_0),
\end{equation}
where $\beta$ is the outer-loop (meta) learning rate. Computing 
$\nabla_{\theta_0}\mathcal{L}_{\mathrm{meta}}$ requires differentiating 
through the $K$ inner-loop update steps in Equation~\eqref{eq7}, which is 
the second-order gradient computation central to MAML. In the context of 
Meta-IFWI, this is implemented by chaining the adjoint-state method through the forward operator $\mathcal{F}$ and standard backpropagation through $N_\theta$.

Once meta-training is complete, Meta-IFWI can be deployed on a previously 
unseen inversion task by warm-starting from $\theta_0$ and performing a 
small number of inner-loop gradient steps following 
Equations~\eqref{eq5}--\eqref{eq7}, without any modification to the 
meta-training procedure. Because $\theta_0$ already encodes shared 
geological priors and inversion patterns learned from the training task 
distribution $p(\mathcal{T})$, far fewer iterations are required to reach a high-quality solution compared with randomly initialized IFWI. The 
advantages of Meta-IFWI in terms of convergence speed, computational 
efficiency, and cross-model generalization are assessed systematically in 
the experiments that follow.

%% file: Sections/Examples.tex
\section{Numerical Examples}\label{sec3}
To evaluate the performance of the proposed Meta-IFWI, we conduct numerical experiments on both in-distribution and out-of-distribution velocity models. All experiments are implemented in PyTorch \citep{paszke2019pytorch} and accelerated on an NVIDIA RTX A6000 GPU. Seismic forward modeling is performed using the Deepwave toolbox \citep{richardson_alan_2026}, and all network parameters are optimized with the Adam optimizer \citep{kingma2014adam}. To ensure a fair comparison, Meta-IFWI and the baseline IFWI share identical network architectures, forward operators, source–receiver configurations, frequency strategies, and optimization hyperparameters. The sole difference lies in initialization: IFWI starts from randomly initialized weights, whereas Meta-IFWI is warm-started from the meta-trained initialization $\theta_0$ described below.

\subsection{Meta-training configuration}
The meta-initialization $\theta_0$ is obtained from a dedicated 
meta-training stage conducted prior to all inversion experiments. A dataset of 36 distinct 2D synthetic velocity models is constructed, covering a wide range of geological scenarios including layered structures, varying velocity gradients, and undulating interfaces. Velocity values across all models are uniformly distributed in the range 1.5–4.8 km/s, encompassing typical sedimentary sequences and complex structural settings. Model spatial scales are varied across tasks to improve the meta-initialization's adaptability to changes in spatial resolution.

At each meta-training iteration, $2T_n$ model–data pairs are randomly sampled from the dataset and partitioned equally into a support set $\mathcal{S}$ (of $T_n$ pairs) and a query set $\mathcal{Q}$ (of $T_n$ pairs), following the procedure described in Method Section. The inner-loop adaptation is driven by $\mathcal{S}$, and the meta-objective is evaluated on the paired tasks in $\mathcal{Q}$. Due to GPU memory limitations, the number of $T_n$ is set to 6 in the following experiments .

The implicit velocity representation network follows a SIREN architecture comprising 6 hidden layers with 128 neurons each. All hidden layers employ the sinusoidal activation $\phi(x) = \sin(\omega_0 x)$, taking spatial coordinates $(x, z)$ as input and outputting P-wave velocity. Meta-training is conducted under the MAML framework with $K = 20$ inner-loop update steps and an inner-loop learning rate of $\alpha = 2\times10^{-3}$. The outer-loop learning rate is initialized at $\beta = 1\times10^{-3}$ and decayed by a factor of 0.8 every 5000 iterations. Meta-training runs for 50000 iterations in total until convergence. Subsequently, the set of meta-trained network parameters $\theta_0$ is used to initialize all inversion experiments. During the IFWI process, the initial learning rate for all experiments is set to $1\times10^{-3}$, and a learning rate decay strategy is adopted to improve convergence stability. The number of iterations for each inversion task is fixed at 500 to ensure a fair comparison.

We test Meta-IFWI on both in-distribution and out-of-distribution models. Performance is assessed through inversion profiles, single-trace velocity comparisons, loss convergence curves, and three quantitative metrics: the mean absolute error (MAE) measuring the average pointwise deviation between the inverted and true velocity models, the structural similarity index 
(SSIM) evaluating the spatial consistency and interface continuity of the reconstructed model, and the signal-to-noise ratio (SNR) quantifying the level of inversion artifacts relative to the true model. Together, these metrics provide a systematic evaluation of Meta-IFWI in terms of inversion accuracy, convergence speed, and generalization capability.

\subsection{In-distribution experiment: Layered model}
We begin with the in-distribution experiments, where the test models share the same geological characteristics as the meta-training distribution. The layered model serves as the first test case.

\begin{figure*}[htbp]
\centering
\includegraphics[height=0.75\textheight,keepaspectratio]{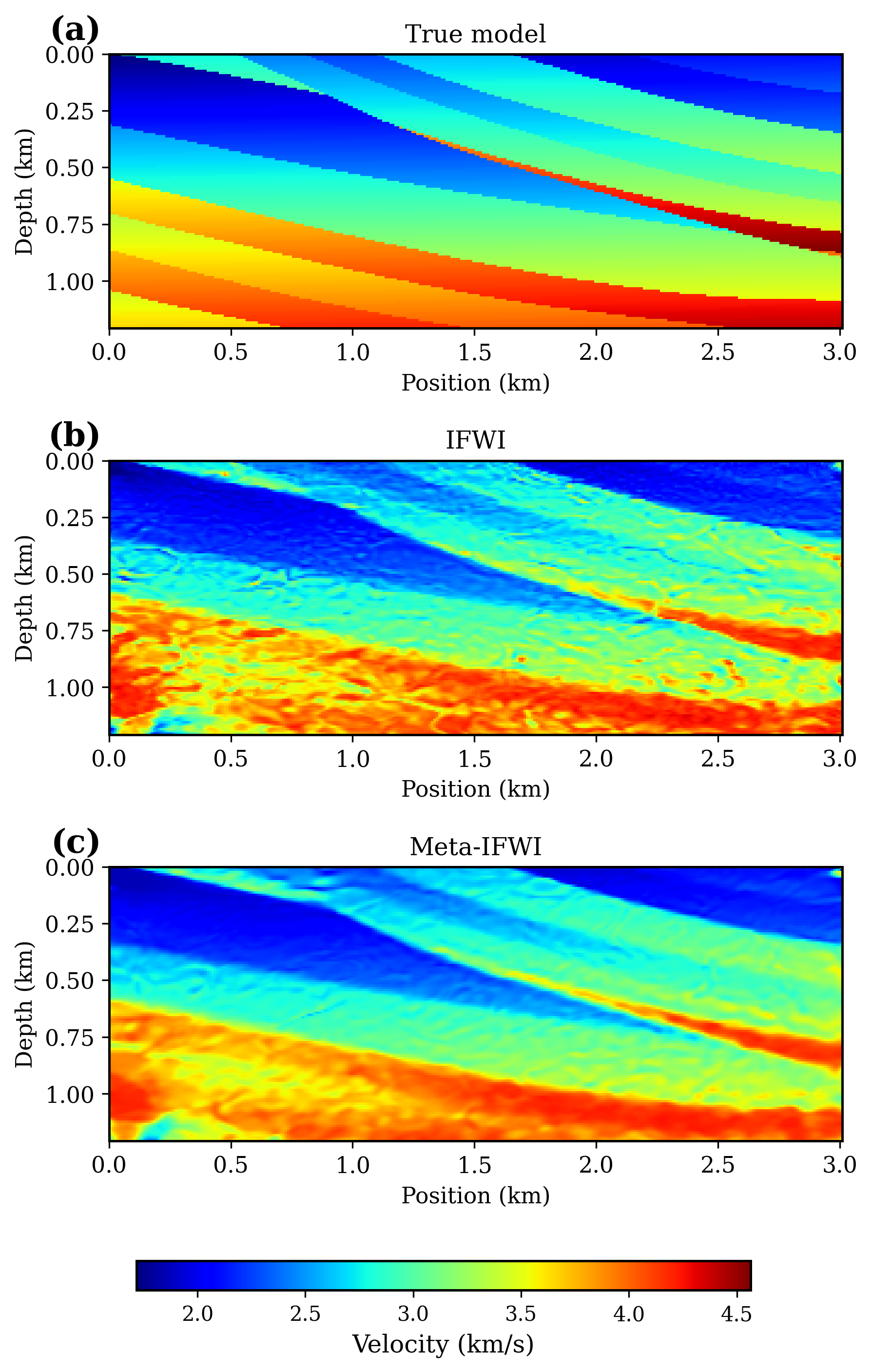}
\caption{Layered model inversion results. (a) represents the true velocity model, (b) shows the IFWI result, and (c) shows the Meta-IFWI result.}
\label{fig1}
\end{figure*}

\begin{figure}[htbp]
\centering
\includegraphics[width=0.75\textwidth]{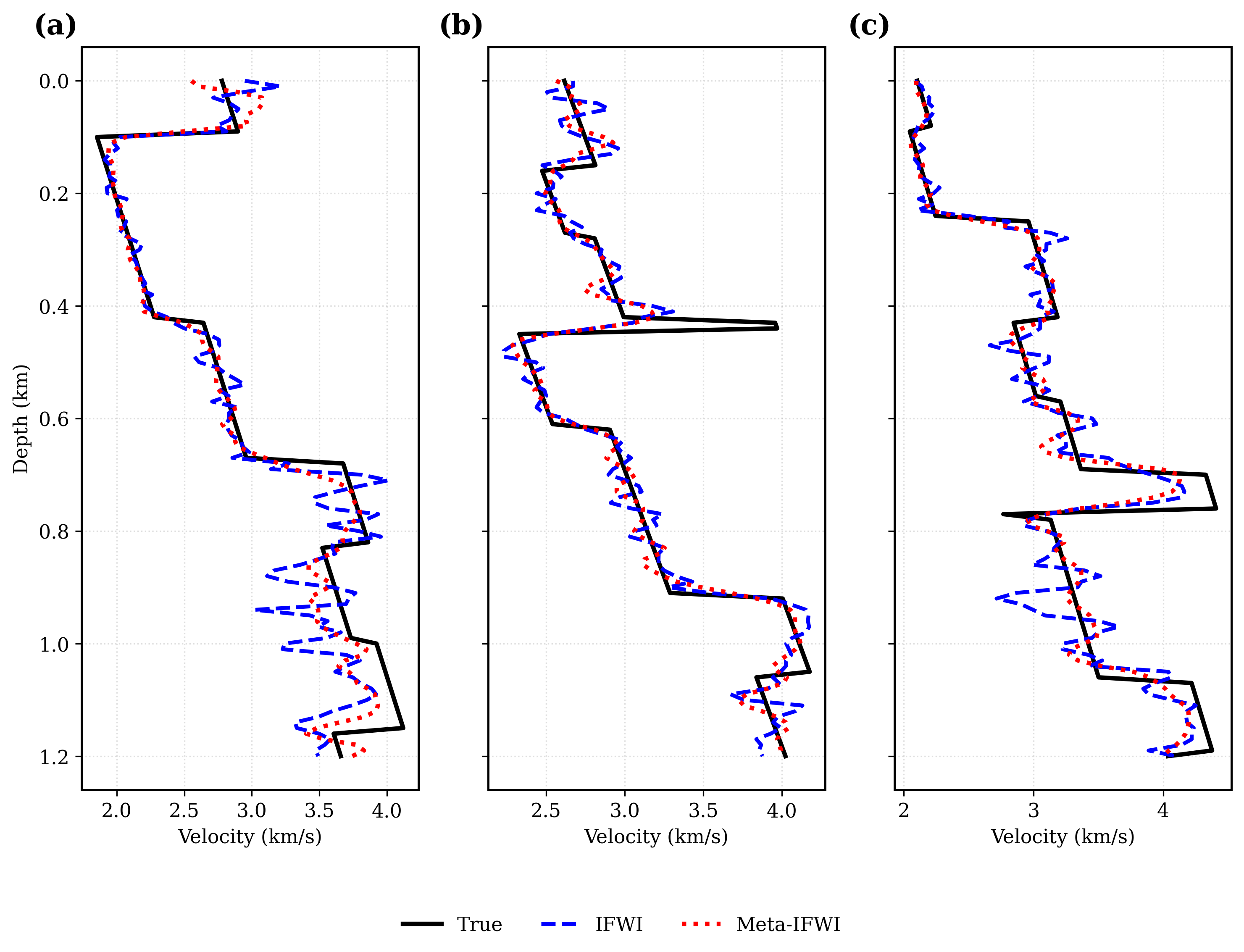}
\caption{Comparison of velocities at (a) 0.5 km, (b) 1.5 km, and (c) 2.5 km in the layered model. The black solid line represents the true velocity, the blue dashed line represents the inverse time-reversal imaging (IFWI) inversion result, and the red dashed line represents the meta-inverse time-reversal imaging (Meta-IFWI) result.}
\label{fig2}
\end{figure}

Figure~\ref{fig1} presents the true velocity model alongside the inversion results of IFWI and Meta-IFWI. Visually, Meta-IFWI more accurately recovers the positions and lateral continuity of the layered interfaces, whereas IFWI exhibits noticeable blurring and misalignment near several interfaces. This difference is further corroborated by the single-trace velocity profiles at horizontal distances of 0.5 km, 1.5 km, and 2.5 km (Figures~\ref{fig2}a-c. Meta-IFWI consistently follows the true model in terms of velocity amplitude, interface depth, and gradient variations, whereas IFWI exhibits noticeable deviations at multiple interfaces.

\begin{figure}[htbp]
\centering
\includegraphics[width=0.75\textwidth]{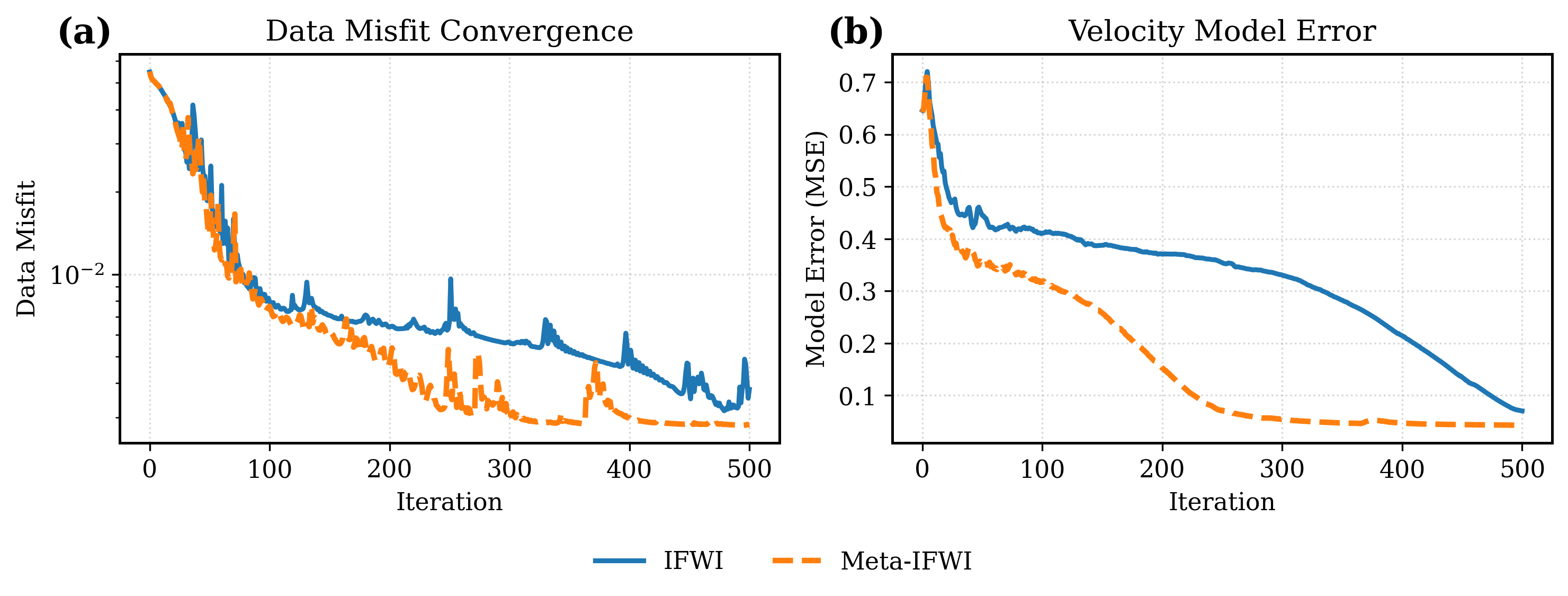}  
\caption{Comparison of convergence curves for the layered model. 
(a) Data misfit convergence and (b) velocity model error. 
The solid blue line represents the IFWI results, whereas the dashed orange line represents the Meta-IFWI results.}
\label{fig3}
\end{figure}

Figure~\ref{fig3} presents the convergence curves for both the data misfit (Figure~\ref{fig3}a) and the velocity model error (Figure~\ref{fig3}b) as functions of iteration. In terms of data misfit, Meta-IFWI achieves a substantially faster reduction and stabilizes at a lower residual, whereas IFWI converges more slowly throughout the optimization. The velocity model error further highlights this difference: Meta-IFWI reaches a stable and low error level by approximately 250 iterations, while IFWI requires the full iterations to achieve a comparable yet still considerably higher error level. This consistent behavior across both metrics is a direct consequence of the meta-learned initialization $\theta_0$. By encoding geological priors from the meta-training distribution, it provides a more favorable starting point in the weight space, allowing the optimization to approach the true solution with fewer gradient updates.

Quantitative metrics (Table~\ref{tab1}) further support these 
observations. Meta-IFWI reduces the MAE from 0.196 km/s to 0.130 km/s, indicating more accurate recovery of absolute velocity values. The SSIM improves from 0.646 to 0.823, reflecting a more faithful reconstruction of spatial continuity across the layered interfaces. The SNR increases by more than 10 dB, confirming effective suppression of inversion artifacts. Together, these results demonstrate that for layered models, which are structurally simple and consistent with the meta-training distribution, Meta-IFWI yields substantial gains in both inversion accuracy and convergence efficiency over randomly initialized IFWI.

\begin{table}[t]
\centering
\caption{Evaluation metrics for layered models.}%
\begin{tabular*}{250pt}{@{\extracolsep\fill}lllll@{\extracolsep\fill}}%
\toprule
\textbf{Method} & \textbf{MAE/(km/s)} & \textbf{SSIM} & \textbf{SNR/(dB)} \\
\midrule
IFWI  & 0.196 & 0.646 & 12.69 \\
Meta-IFWI  & 0.130 & 0.823 & 23.05 \\
\bottomrule
\end{tabular*}
\label{tab1}
\end{table}

\subsection{In-distribution experiment: Overthrust model}
The second in-distribution test considers the Overthrust model, which presents a significantly more challenging scenario characterized by pronounced faults.

\begin{figure*}[htbp]
\centering
\includegraphics[height=0.75\textheight,keepaspectratio]{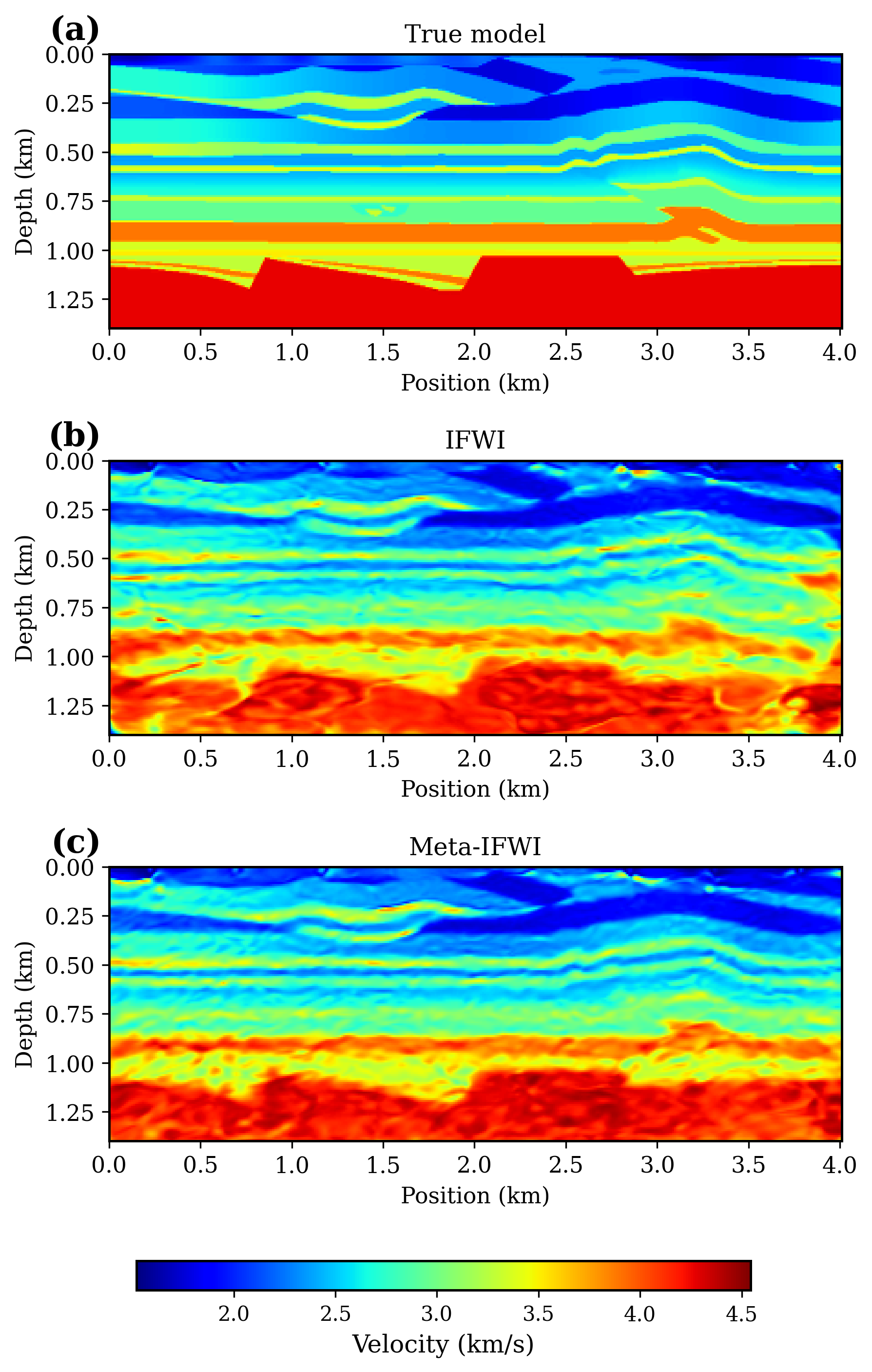}
\caption{Overthrust model inversion results.(a) represents the true velocity model, (b) shows the IFWI result, and (c) shows the Meta-IFWI result.}
\label{fig4}
\end{figure*}

\begin{figure}[htbp]
\centering
\includegraphics[width=0.75\textwidth]{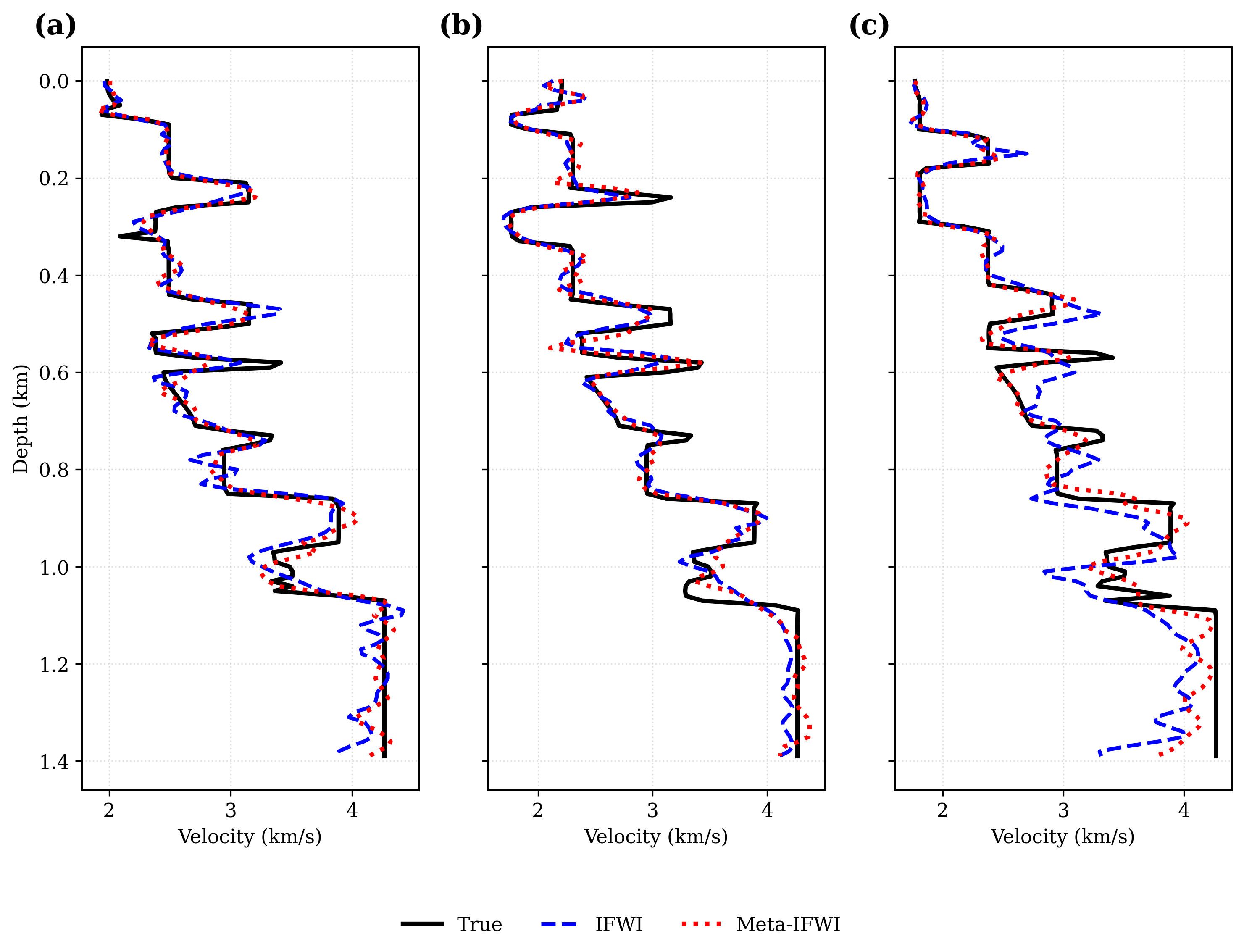}  
\caption{Comparison of velocities at (a) 1.0 km, (b) 2.0 km, and (c) 3.5 km in the Overthrust model. The black solid line represents the true velocity, the blue dashed line represents the inverse time-reversal imaging (IFWI) inversion result, and the red dashed line represents the meta-inverse time-reversal imaging (Meta-IFWI) result.}
\label{fig5}
\end{figure}

Figure~\ref{fig4} presents the true velocity model alongside the inversion results of IFWI and Meta-IFWI. Meta-IFWI successfully recovers the continuity of major velocity interfaces and fault structures even in regions of high structural complexity, whereas IFWI produces noticeable velocity distortions and structural blurring where faulting is dense and velocity variations are strong. The single-trace profiles at horizontal distances of 1.0 km, 2.0 km, and 3.5 km (Figures~\ref{fig5}a-c) further highlight this contrast: Meta-IFWI closely matches the true model in interface depth, velocity amplitude, and gradient variations, whereas IFWI exhibits systematic shifts and velocity errors near multiple interfaces.

\begin{figure}[htbp]
\centering
\includegraphics[width=0.75\textwidth]{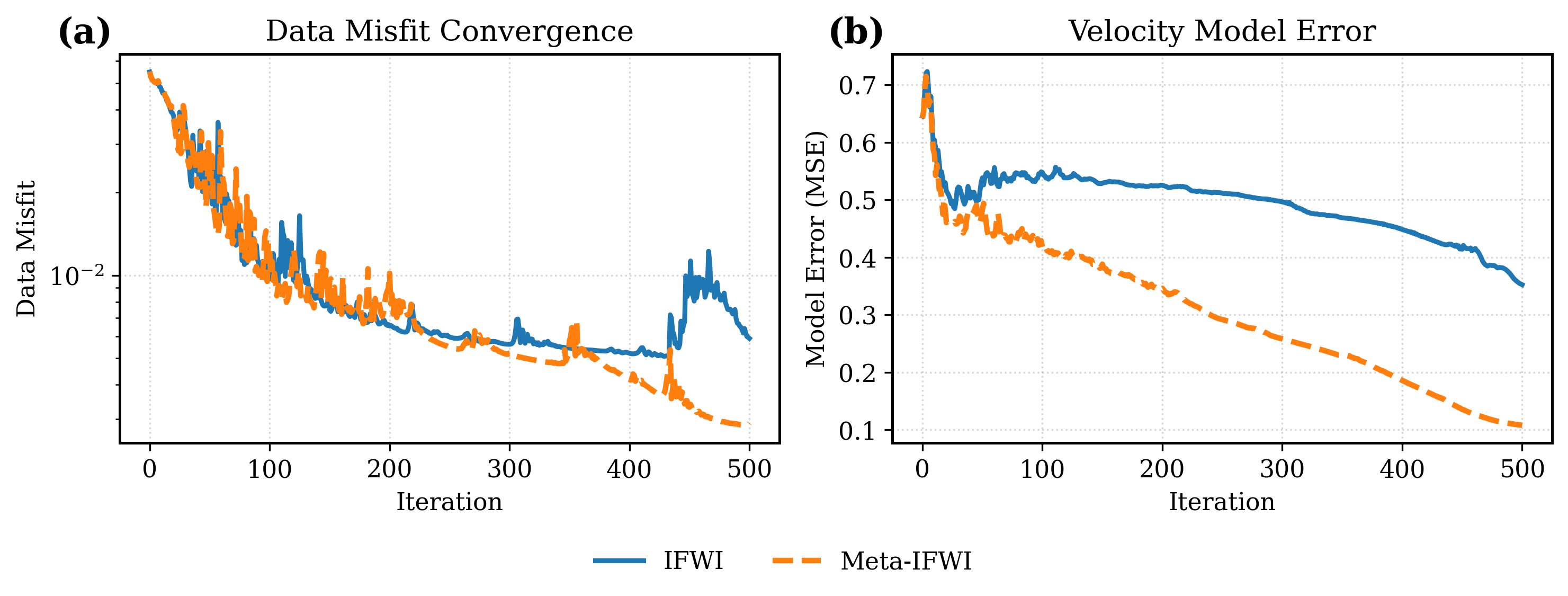}  
\caption{Comparison of convergence curves for the Overthrust model. 
(a) Data misfit convergence and (b) velocity model error. 
The solid blue line represents the IFWI results, whereas the dashed orange line represents the Meta-IFWI results.}
\label{fig6}
\end{figure}

Figure~\ref{fig6} presents the convergence curves for the data misfit (Figure~\ref{fig6}a) and the velocity model error (Figure~\ref{fig6}b) for the Overthrust model. In the data misfit, both methods decrease at a comparable rate during the early iterations. However, Meta-IFWI progressively pulls ahead from around iteration 200 onward and reaches a substantially lower residual by the end of the optimization, while IFWI 
exhibits a prominent spike near iteration 450 and fails to sustain a consistent downward trend. The velocity model error reveals a more fundamental difference in optimization behavior. IFWI decreases rapidly in the first 50 iterations but then remains at a high error level for the remainder of the run, 
indicating that the IFWI is trapped in a poor local minimum. Meta-IFWI avoids this stagnation entirely, maintaining a steady and continuous decrease throughout all 500 iterations. This pattern is considerably more pronounced than what was observed for the layered model, consistent with the view that the benefit of meta-learned initialization grows with structural complexity.

Quantitative metrics in Table~\ref{tab2} confirm these observations. Meta-IFWI reduces the MAE from 0.383 km/s to 0.187 km/s (a reduction of more than 50\%), which indicates a substantial gain in velocity accuracy under complex structural conditions. The SSIM improves from 0.596 to 0.792, reflecting better preservation of structural consistency across faulted and heterogeneous regions. The SNR increases from 8.40 dB to 17.05 dB, demonstrating markedly enhanced suppression of inversion artifacts. Notably, the performance gains on the Overthrust model exceed those observed for the layered model across all three metrics, underscoring that meta-learned initialization plays an increasingly critical role as geological complexity grows.

\begin{table}[htbp]
\centering
\caption{Evaluation metrics for Overthrust model.}%
\begin{tabular*}{250pt}{@{\extracolsep\fill}lllll@{\extracolsep\fill}}%
\toprule
\textbf{Method} & \textbf{MAE/(km/s)} & \textbf{SSIM} & \textbf{SNR/(dB)} \\
\midrule
IFWI  & 0.383 & 0.596 & 8.40 \\
Meta-IFWI  & 0.187 & 0.792 & 17.05 \\
\bottomrule
\end{tabular*}
\label{tab2}
\end{table}

\subsection{Out-of-distribution experiment: Marmousi 2 model}
The final experiment evaluates the generalization capability of Meta-IFWI on an out-of-distribution model. The Marmousi 2 model, which is one of the most classic models in the seismic exploration community, features significantly more complex velocity variations and lateral heterogeneity, making it a strict test of whether the meta-learned initialization $\theta_0$ can transfer beyond its training distribution.

\begin{figure*}[htbp]
\centering
\includegraphics[height=0.75\textheight,keepaspectratio]{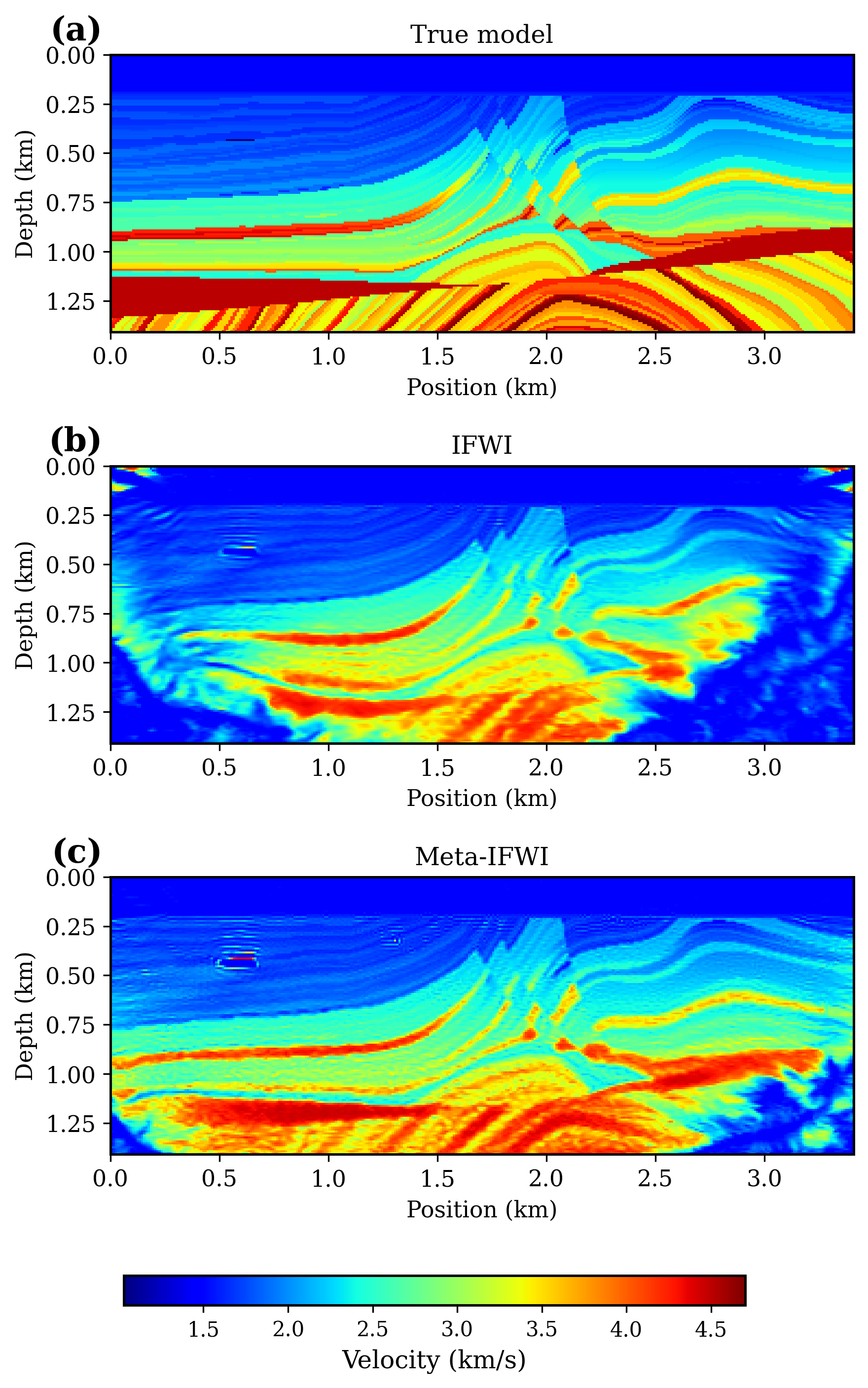}
\caption{Marmousi 2 model inversion results.(a) represents the true velocity model, (b) shows the IFWI result, and (c) shows the Meta-IFWI result.}
\label{fig7}
\end{figure*}

\begin{figure}[htbp]
\centering
\includegraphics[width=0.75\textwidth]{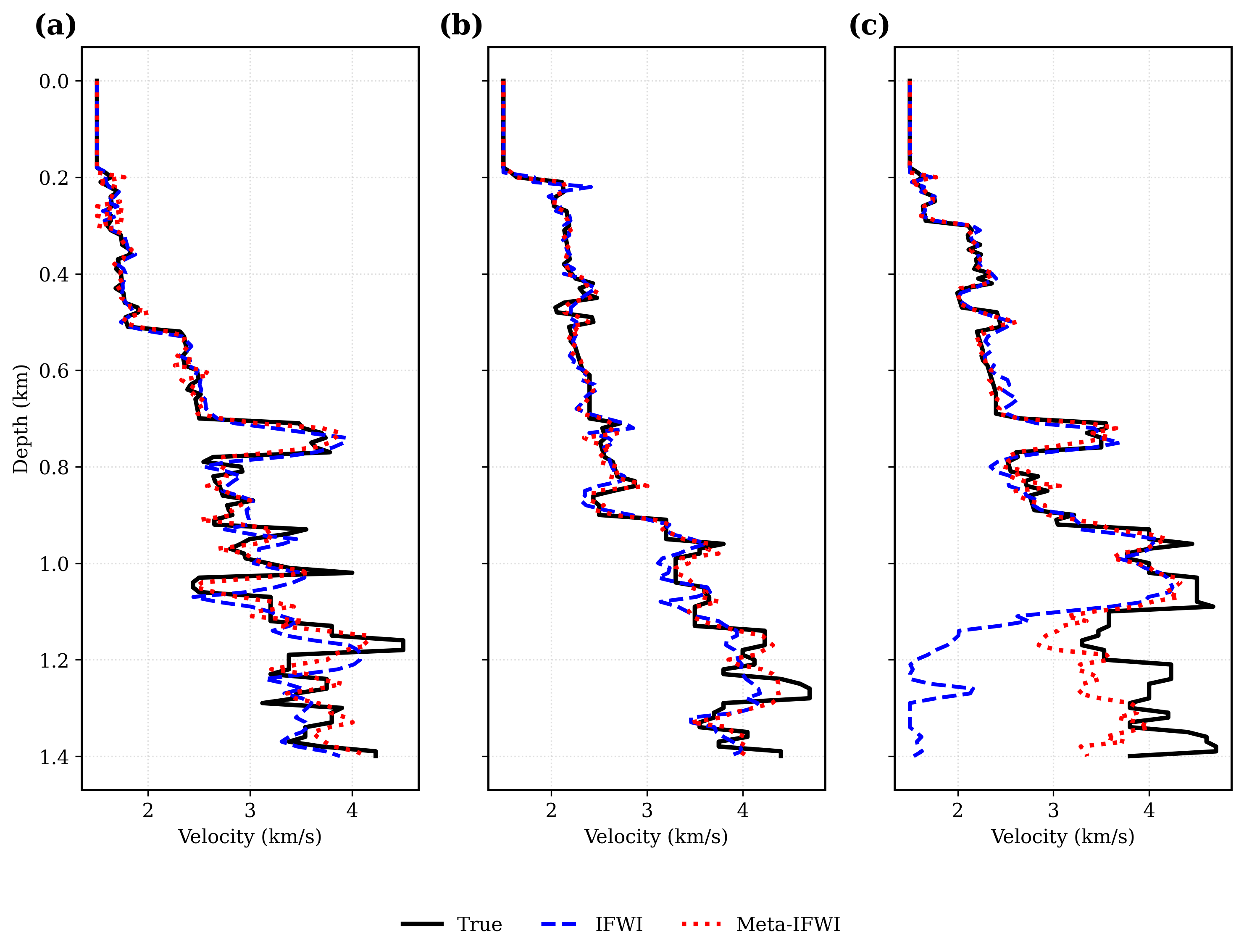}  
\caption{Comparison of velocities at (a) 1.5 km, (b) 2.0 km, and (c) 2.5 km in the Marmousi 2 model. The black solid line represents the true velocity, the blue dashed line represents the inverse time-reversal imaging (IFWI) inversion result, and the red dashed line represents the meta-inverse time-reversal imaging (Meta-IFWI) result.}
\label{fig8}
\end{figure}

Figure \ref{fig7} presents the inversion results for the Marmousi 2 model, where (a) denotes the true velocity model, (b) is the IFWI result, and (c) is the Meta-IFWI result. Despite the distributional shift, Meta-IFWI constructs a reasonable velocity background and recovers the main geological structures, including faults, low-velocity zones, and high-velocity slopes, with smooth and laterally continuous interfaces. IFWI, by contrast, produces velocity distortions and structural blurring in regions of dense faulting and steep velocity gradients, with noticeable interface misalignments. The single-trace profiles at horizontal distances of 1.5 km, 2.0 km, and 2.5 km (Figure~\ref{fig8}a-c) further reinforce this contrast: Meta-IFWI accurately captures velocity trends in both shallow and mid-to-deep layers, closely matching the true model in interface depth and velocity values, whereas IFWI shows systematic underestimation or overestimation at multiple interfaces.

\begin{figure}[htbp]
\centering
\includegraphics[width=0.75\textwidth]{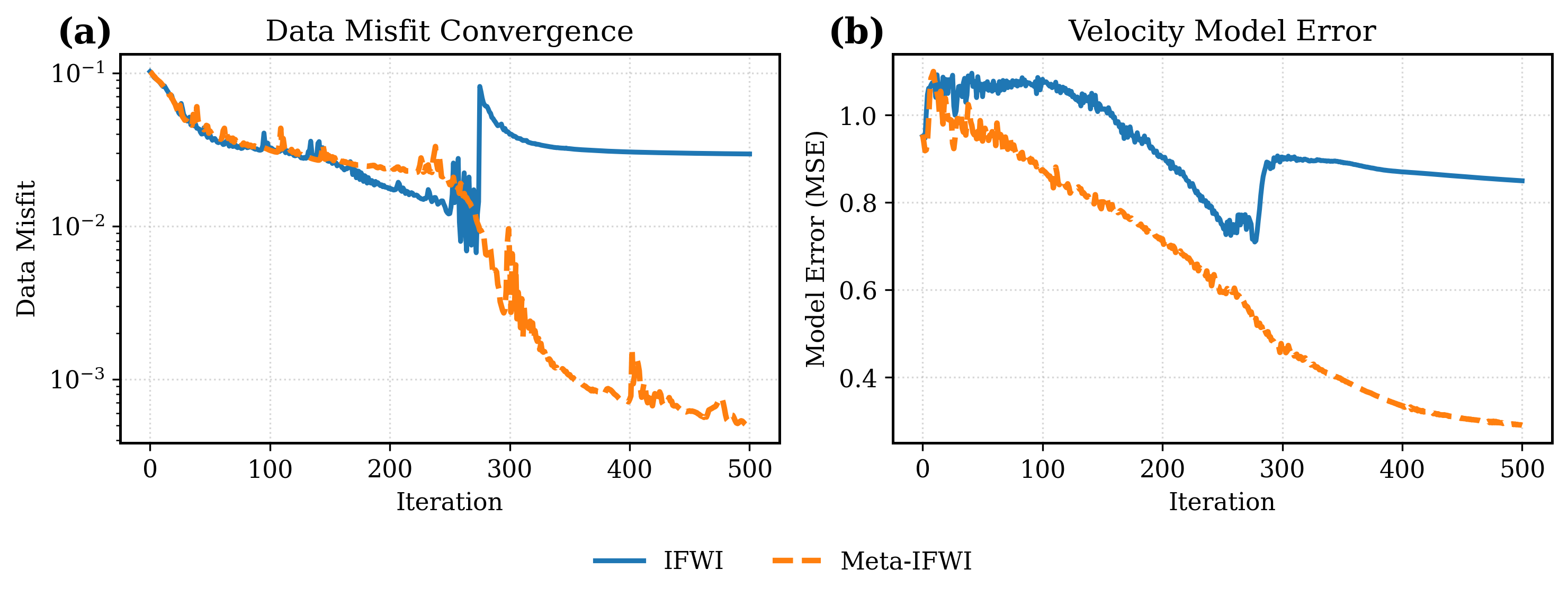}  
\caption{Comparison of convergence curves for the Marmousi 2 model. 
(a) Data misfit convergence and (b) velocity model error. 
The solid blue line represents the IFWI results, whereas the dashed orange line represents the Meta-IFWI results.}
\label{fig9}
\end{figure}

Figure~\ref{fig9} presents the convergence curves for the Marmousi2 model. In the data misfit (Figure~\ref{fig9}a), both methods follow a similar decreasing trend during the first 250 iterations, after which IFWI suffers a sharp spike and subsequently flattens at a high residual level, suggesting the optimization has stalled. Meta-IFWI, by contrast, undergoes a rapid and sustained reduction from around iteration 270 onward, reaching a final residual nearly two orders of magnitude lower. For velocity model error (Figure~\ref{fig9}b), IFWI remains near its initial error level for most of the run and shows no meaningful recovery after the transition point, while Meta-IFWI continues to improve steadily throughout all 500 iterations. The degree of stagnation seen here is notably more severe than in the in-distribution experiments, highlighting that a well-informed initialization becomes especially important when the target geology lies outside the training distribution.

Quantitative metrics in Table~\ref{tab3} confirm these observations. Meta-IFWI reduces the MAE from 0.376 km/s to 0.272 km/s (a reduction of over 30\%), improves the SSIM from 0.594 to 0.748, and increases the SNR by more than 8 dB. While the absolute gains are somewhat smaller than those observed for the in-distribution Overthrust model, which is expected given the distributional shift, Meta-IFWI maintains consistently higher accuracy and structural fidelity than IFWI across all three metrics, demonstrating robust generalization of the meta-learned initialization to previously unseen geological settings.

In summary, Meta-IFWI not only performs well on in-distribution tasks but also achieves significant improvements in accuracy and stability on out-of-distribution complex models, providing a reliable initialization and generalization capability for implicit full-waveform inversion.

\begin{table}[htbp]
\centering
\caption{Evaluation metrics for Marmousi 2 model.}%
\begin{tabular*}{250pt}{@{\extracolsep\fill}lllll@{\extracolsep\fill}}%
\toprule
\textbf{Method} & \textbf{MAE/(km/s)} & \textbf{SSIM} & \textbf{SNR/(dB)} \\
\midrule
IFWI  & 0.376 & 0.594 & 10.23 \\
Meta-IFWI  & 0.272 & 0.748 & 16.59 \\
\bottomrule
\end{tabular*}
\label{tab3}
\end{table}

%% file: Sections/Discussions.tex
\section{Discussion}

The three experiments presented above collectively reveal a consistent and systematic pattern: the performance gains of Meta-IFWI over IFWI grow with increasing geological complexity. For the relatively simple layered model, Meta-IFWI already achieves notable improvements across all evaluation metrics. These gains become more pronounced for the structurally complex Overthrust model, and Meta-IFWI maintains a consistent advantage even under the distributional shift of the Marmousi 2 model. This progressive pattern points to a unified underlying mechanism: in the high-dimensional, non-convex weight space of the implicit neural network, the meta-learned initialization $\theta_0$ places the optimization in the vicinity of a geologically meaningful subspace from the very first iteration, effectively shortening the optimization path. As geological complexity increases and the misfit landscape becomes more non-convex, this favorable starting point becomes increasingly critical and, therefore, explains why the advantage of Meta-IFWI is most pronounced precisely where conventional IFWI struggles most.

The Marmousi 2 results deserve particular attention, as they test a capability that is not guaranteed by the meta-training design: generalization to geological structures not encountered during training. Despite the significant distributional shift, Meta-IFWI maintains consistent superiority over IFWI across all evaluation metrics, suggesting that $\theta_0$ does not merely memorize specific geological patterns from the training set. Instead, it encodes more general inversion priors (such as smooth velocity gradients, laterally continuous interfaces, and physically plausible velocity ranges) that transfer across a broad class of geological settings. While the absolute gains on Marmousi 2 are somewhat smaller than those on the Overthrust model, as expected given the increased distributional distance, the consistent advantage of Meta-IFWI confirms robust generalization. We attribute this behavior to the intentional design of the meta-training dataset, which spans a wide range of velocity values, structural complexities, and spatial scales, encouraging $\theta_0$ to capture task-agnostic inversion patterns rather than task-specific features.

Beyond accuracy, the convergence curves observed across all three experiments provide further mechanistic insight into why Meta-IFWI performs better. In all cases, Meta-IFWI achieves a steeper misfit reduction in the early iterations and stabilizes at a lower residual, while IFWI converges more slowly and exhibits greater susceptibility to local fluctuations, an effect that intensifies from the layered model to the Overthrust and Marmousi 2 models as the nonlinearity of the misfit landscape increases. This behavior is a direct consequence of the difference in initialization: randomly initialized IFWI must first traverse a large, geologically uninformative region of the parameter space before approaching a plausible solution, whereas Meta-IFWI begins its search already within a physically meaningful neighborhood. The result is not only faster convergence but also improved optimization stability, which ultimately translates into the more accurate and structurally faithful inversion profiles observed across 
all experiments.

Despite these encouraging results, several limitations of the current work should be acknowledged. The meta-training dataset comprises only 36 synthetic 2D velocity models, and while this is sufficient to demonstrate strong in-distribution performance and moderate out-of-distribution generalization, the meta-initialization may degrade for geological settings that are extreme outliers relative to the training distribution. Furthermore, the second-order gradient computation required by MAML imposes a GPU memory overhead that constrains the number of task pairs per iteration, and all experiments are conducted under the acoustic wave assumption, leaving the extension to elastic or multiparameter inversion as an open problem. Finally, the framework has only been validated on synthetic data. Its behavior on field seismic data, where noise characteristics and source uncertainties are more severe, remains to be investigated.

These limitations point naturally to several promising directions for future work. Expanding the meta-training dataset in both size and geological diversity is expected to further improve out-of-distribution generalization. Replacing full second-order MAML with first-order approximations such as Reptile \citep{nichol2018reptile} could substantially reduce memory consumption, enabling larger task batches and potentially improving meta-training stability. Extending Meta-IFWI to elastic FWI and multiparameter joint inversion represents another important direction, as the implicit neural representation is naturally suited to simultaneously encoding multiple physical parameters. Finally, applying Meta-IFWI to field datasets in conjunction with frequency-progressive inversion strategies would provide the most direct assessment of its practical utility.

%% file: Sections/Conclusions.tex
\section{Conclusions}\label{sec5}
We proposed a Meta-Learning based implicit full waveform inversion (Meta-IFWI) method, which integrated implicit neural representations, wave equation forward modeling constraints, and a meta-learning framework to efficiently and stably solve seismic inverse problems. This approach represents subsurface physical parameters through continuous implicit functions, maintaining high resolution and differentiability while avoiding the strong dependence on grid discretization for resolution and memory resources that traditional methods entail. On this basis, the implicit full waveform inversion problem is formulated as a meta-learning task. By learning shared initialization parameters across multiple inversion tasks, the model can rapidly converge on new seismic observations with only a few gradient updates. Numerical experiments demonstrated that, compared with conventional IFWI starting from random initialization, Meta-IFWI significantly reduces the number of inversion iterations, accelerates convergence, and exhibits more stable inversion performance and stronger generalization on both in-distribution and out-of-distribution models with complex geological structures. 


%% file: Sections/Acknowledgement.tex
\section{Acknowledgments}
This work was supported by the National Natural Science Foundation of China under Grant Nos. 42130808 and 42574181. The author Shijun Cheng thanks King Abdullah University of Science and Technology for supporting this research.